\documentclass[sigconf, screen]{acmart}
\usepackage{xspace}
\usepackage{glossaries}
\usepackage{amsmath}
\usepackage{mathtools}
\usepackage{wrapfig}
\usepackage{url}
\usepackage{xcolor, colortbl}
\definecolor{acmgreen}{RGB}{0,128,0}

\AtBeginDocument{%
  }

\setcopyright{rightsretained}
\copyrightyear{2025}
\acmYear{2025}
\acmConference{SIGGRAPH Posters '25}{August 10-14, 2025}{Vancouver, BC, Canada}\acmBooktitle{Special Interest Group on Computer Graphics and Interactive Techniques Conference Posters (SIGGRAPH Posters '25), August 10-14, 2025}\acmDOI{10.1145/3721250.3742993}
\acmISBN{979-8-4007-1549-5/2025/08}

\newacronym{VAE}{VAE}{Variational Auto-Encoder}
\newacronym{INR}{INR}{Implicit Neural Representations}
\newacronym{PSNR}{PSNR}{Peak Signal-to-Noise Ratio}
\newacronym{SSIM}{SSIM}{Structural Similarity Index Measure}
\newacronym{3D}{3D}{Three-Dimensional}
\newacronym{MLP}{MLP}{Multi-Layer Perceptron}
\newacronym{CGH}{CGH}{Computer-Generated Holography}
\newacronym{CNN}{CNN}{Convolutional Neural Network}

\global\long\def\VAE{\acrshort{VAE}\xspace}
\global\long\def\INR{\acrshort{INR}\xspace}

\global\long\def\TAESD{\textcolor{ColorTAESD}{\textbf{TAESD}}\xspace}

\global\long\def\vanillaMLP{\textcolor{ColorMLP}{\textbf{vanilla MLP}}\xspace}
\global\long\def\SIREN{\textcolor{ColorSIREN}{\textbf{SIREN}}\xspace}
\global\long\def\FILMSIREN{\textcolor{ColorFILMSIREN}{\textbf{FilmSIREN}}\xspace}

\newcommand{\refFig}[1]{Fig.~\ref{fig:#1}}

\newcommand{\refTbl}[1]{Tbl.~\ref{tbl:#1}}

\definecolor{ColorMLP}{rgb}{0.8470588235294118, 0.10588235294117647, 0.3764705882352941}
\definecolor{ColorSIREN}{rgb}{0.11764705882352941, 0.5333333333333333, 0.8980392156862745}
\definecolor{ColorFILMSIREN}{rgb}{1, 0.7568627450980392, 0.027450980392156862}
\definecolor{ColorTAESD}{rgb}{0, 0.30196078431372547, 0.25098039215686274}

\newcommand{\phaseonlyhologram}{P\xspace}
\newcommand{\wavelengths}{\lambda\xspace}
\newcommand{\pixelpitch}{px\xspace}
\newcommand{\bottleneck}{\alpha\xspace}

\begin{document}

\title{Assessing Learned Models for Phase-only Hologram Compression}

\author{Zicong Peng}
\affiliation{%
  \institution{University College London}
  \city{London}
  \country{UK}
}
\email{zicong.peng.24@ucl.ac.uk}

\author{Yicheng Zhan}
\affiliation{%
  \institution{University College London}
  \city{London}
  \country{UK}
}
\email{ucaby83@ucl.ac.uk}

\author{Josef Spjut}
\affiliation{%
  \institution{NVIDIA}
  \city{Durham, NC}
  \country{US}
}
\email{jspjut@nvidia.com}

\author{Kaan Akşit}
\affiliation{%
  \institution{University College London}
  \city{London}
  \country{UK}
}
\email{k.aksit@ucl.ac.uk}

\renewcommand{\shortauthors}{Peng et al.}

\begin{abstract}
  We evaluate the performance of four common learned models utilizing \INR and \VAE structures for compressing phase-only holograms in holographic displays.  
  The evaluated models include a \vanillaMLP, \SIREN \cite{sitzmann2020implicit}, and \FILMSIREN \cite{chan2021pi}, with \TAESD \cite{bohan2023tiny} as the representative \VAE model.
  Our experiments reveal that a pretrained image \VAE, \TAESD, with $2.2M$ parameters struggles with phase-only hologram compression, revealing the need for task-specific adaptations.
  Among the \INR{}s, \SIREN with $4.9k$ parameters achieves $\%40$ compression with high quality in the reconstructed 3D images (PSNR = $34.54$~dB).
  These results emphasize the effectiveness of \INR{}s and identify the limitations of pretrained image compression \VAE{}s for hologram compression task.
\end{abstract}

\begin{CCSXML}
<ccs2012>
   <concept>
       <concept_id>10010147.10010371.10010395</concept_id>
       <concept_desc>Computing methodologies~Image compression</concept_desc>
       <concept_significance>500</concept_significance>
       </concept>
   <concept>
       <concept_id>10003752.10003809.10010031.10002975</concept_id>
       <concept_desc>Theory of computation~Data compression</concept_desc>
       <concept_significance>500</concept_significance>
       </concept>
   <concept>
       <concept_id>10010583.10010588.10010591</concept_id>
       <concept_desc>Hardware~Displays and imagers</concept_desc>
       <concept_significance>300</concept_significance>
       </concept>
   <concept>
       <concept_id>10010583.10010786.10010810</concept_id>
       <concept_desc>Hardware~Emerging optical and photonic technologies</concept_desc>
       <concept_significance>300</concept_significance>
       </concept>
 </ccs2012>
\end{CCSXML}

\ccsdesc[500]{Computing methodologies~Image compression}
\ccsdesc[500]{Theory of computation~Data compression}
\ccsdesc[300]{Hardware~Displays and imagers}
\ccsdesc[300]{Hardware~Emerging optical and photonic technologies}

\keywords{Computer-Generated Holography, Hologram Compression, Holographic Displays}
\begin{teaserfigure}
  \includegraphics[width=\textwidth]{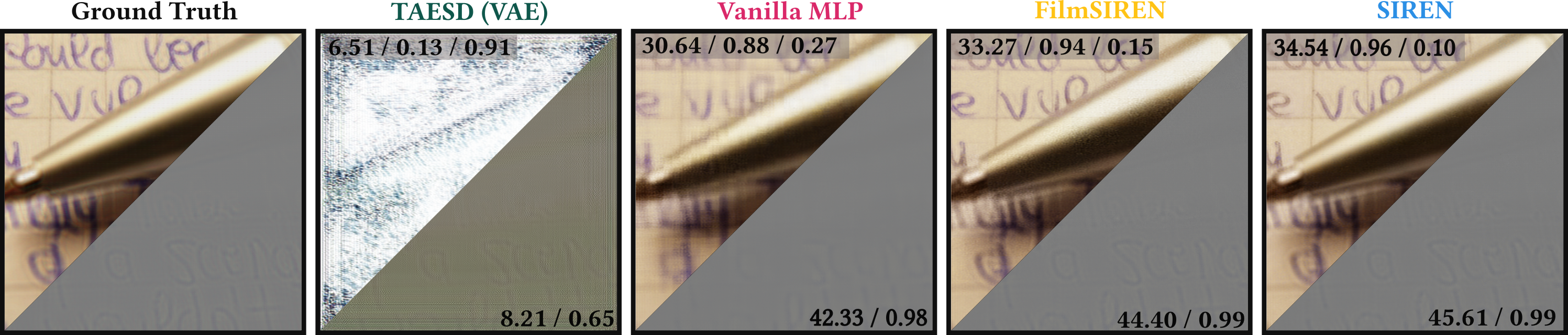}
  \caption{
    Comparing common \INR and \VAE based learned models for the hologram compression task in a split view.
    For every image inset, the lower triangles represent the compressed phase-only hologram with PSNR and SSIM metrics, respectively.
    The upper triangle displays the 3D reconstruction of the corresponding 3D phase-only hologram, incorporating PSNR, SSIM ,and LPIPS metrics in order
    (Source image: \href{https://openverse.org/image/df8f4a77-6b7f-434f-a38a-abe5cf1f4f79?q=text&p=1}{photosteve101}).
          }
  \Description{}
  \label{fig:teaser}
\end{teaserfigure}


\maketitle

\section{Introduction}
\begingroup
\setlength{\intextsep}{0.5pt}
\setlength{\columnsep}{10pt}
\begin{wrapfigure}{r}{0.35\columnwidth}
\centering
\includegraphics[width=1.0\linewidth]{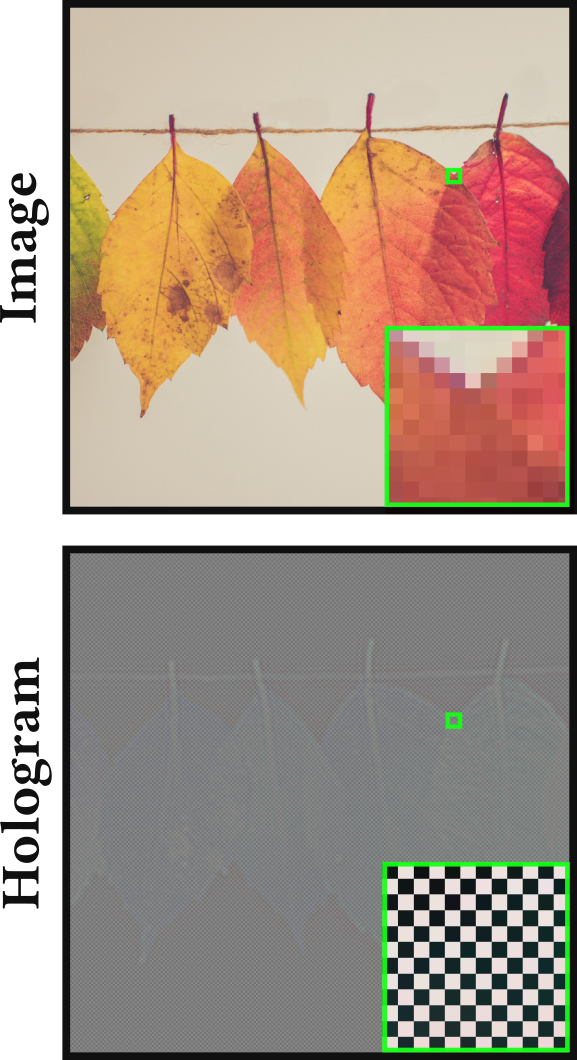}
\caption{
Unlike natural images, holograms are dominated by high-frequency content.
(Source image: \href{https://openverse.org/image/15c29763-6088-432a-8246-9475f1672eb2?q=colorful&p=14}{leavesColor2025})
}
\label{fig:high_frequency}
\end{wrapfigure}
Emerging holographic displays \cite{kavakli2023multicolor, zheng2024focalholography} utilize phase-only holograms to reconstruct full-color 3D scenes at various optical depths.
The diversity of phase-only hologram types and encoding schemes available today reflects the infancy of display hardware and the lack of standardized practices.
These hologram types, which include single-color and multi-color variants, and their counterpart encoding schemes like double-phase or direct phase encoding, 
require efficient compression due to their high-frequency information content (see \refFig{high_frequency}).
% TAESD is specifically designed for natural images - it compresses by discarding the high-frequency details that are not sensitive to the human eye. However, the high-frequency structure of the hologram directly determines the quality of 3D reconstruction. 
% In contrast, during INR compression, the key information can be remained, thereby avoiding artifacts or distortion.
On the other hand, learned models \cite{wang2022joint} are known to struggle preserving high-frequency features, and the potential benefits of compression using learned models remain uncertain, regardless of type or encoding.
This ambiguity motivates our investigation into whether learned models can effectively compress phase-only holograms and contribute to improved storage and transmission efficiency.
Therefore, we chose two general learned model structures: \VAE, which generates images by learning in a latent space, and \INR, which directly represents image content using implicit neural functions.
Our work presents the first attempt to assess the learned models in a structured way for the hologram compression task. 
We focus solely on comparing the performance of different existing methods for hologram image compression with the aim of providing a foundation for future optimization using novel models or approaches. 
Specifically, in this early study, we analyze a \vanillaMLP, \SIREN \cite{sitzmann2020implicit}, and \FILMSIREN \cite{chan2021pi}, with \TAESD \cite{bohan2023tiny} as the representative \VAE model.

\section{Method}
Our assessments involve using single-color double-phase encoded phase-only holograms, $\phaseonlyhologram \in \mathbb{R}^{3 \times 512 \times 512}$, using three color primaries.
These $\phaseonlyhologram$s are calculated for three wavelengths, $\wavelengths = \{473, 515, 639\}\,nm$ and a fixed pixel pitch, $\pixelpitch = 3.74\,\mu m$ (Jasper Display JD7714).
We adopt an off-the-shelf \TAESD trained for image compression task.
Specifically, the \TAESD with $2.2M$ parameters encodes $\phaseonlyhologram$ to a $\bottleneck \in \mathbb{R}^{16 \times 64 \times 64}$ and later decodes into the original resolution of $3 \times 512 \times 512$.
% As observed in \refFig{teaser}, a pretrained \TAESD fails to encode and decode the provided $\phaseonlyhologram$s.
\refFig{teaser} shows pretrained \TAESD fails, requiring dedicated training for generalization,
% \TAESD or any other \VAE would require a dedicated training to generalize over a variety of $\phaseonlyhologram$s.
% In addition, when the feature size of $\bottleneck$ and $\phaseonlyhologram$ is compared, the compression would only amount to $\%92$ reduction in feature size, excluding the parameter count of \TAESD.
the compression would amount to $\%92$ reduction (excluding \TAESD params).
Thus, we choose to explore \INR based models to see if the feature size could be further reduced while accepting longer training times as \INR{}s typically are overfitted on a single data at a time.
In our study, we compare three foundational \INR architectures (\vanillaMLP, \SIREN, and \FILMSIREN), aiming for $\sim \%40$ compression to strike a balance between the quality of the reconstructed image and the compression ratio. 
% We divide given $\phaseonlyhologram${}s into patches with various sizes in our experiments (\ie $3 \times 64 \times 64$ or $3 \times 128 \times 128$).
% We train a model for each patch, and we combine these models together to reconstruct the entire hologram at original resolution.
% In our training, we initialize each training with the weights learned from the previous patch to avoid weights leading to visually different hologram patches.%
$\phaseonlyhologram$s are split into patches (e.g., $3 \times 64 \times 64$), a separate model is trained for each patch (initialized from prior weights), and their outputs are combined for full reconstruction.
Experiments that we are going to detail in the next section utilize ten different holograms (The purpose of selecting a small but diverse set of initial samples in this study is to demonstrate the comparative trends among different methods. The large-scale validation work will be addressed in the subsequent research.) and turns them into patches by following the choices listed in \refTbl{comparison}.
% In all our trainings with \INR based models, we used an Adam optimizer with a learning rate of $0.0001$and gradually decreased the learning rate using StepLR scheduler, which reduces the learning rate by a factor of $0.5$ every $5000$ epochs.
% We train all \INR based models for $5000$ epochs.
All \INR{}s use Adam (lr=0.0001) with StepLR (gamma=0.5 every 5000 epochs), trained for 10000 epochs.
More details of our models are available in our supplementary.

\section{Results and Discussions}
% Our study reveals that \SIREN and \FILMSIREN offer good compression capabilities, with \vanillaMLP showing less competitive results. 
% Among the tested architectures, \SIREN demonstrates better consistency across varying patch configurations. 
\SIREN and \FILMSIREN provide strong compression, outperforming \vanillaMLP, with \SIREN showing best consistency.
In our current experiments in \refTbl{comparison}, \SIREN achieves the highest fidelity with a PSNR of 42.29 dB and SSIM of 0.9997 at $3 \times 64 \times 64$ patch size, 
For 3D holographic reconstructions with a $5$ mm volume depth, metrics are averaged over three focal planes at propagation distances of $-2.5$, $0$, and $+2.5$ mm.
Under this setting, \SIREN attains a PSNR of 34.54 dB, SSIM of 0.96, LPIPS of 0.10 marginally outperforms \FILMSIREN (PSNR = 33.27 dB, SSIM = 0.94, LPIPS = 0.15). 
Additionally, under identical training schedules, both \SIREN and \FILMSIREN frequently satisfied the early stopping criterion near 2000 epochs. 
This consistency implies a relatively smooth optimization process, suggesting that these models can converge effectively without compromising image quality, 
which is a favorable property in hologram compression task.
% The computational demands—approximately $T$ hours per hologram—represent a trade-off that appears justified by \SIREN's quality preservation capabilities. 
The computational demand of around $40$ minutes per hologram seems justified by \SIREN's ability to preserve quality.
These observations suggest that specialized \INR architectures require further investigation for the hologram compression task, 
potentially opening new solutions for efficient 3D scene representation in holographic displays.
% Compressing phase-only holograms in a robust manner is an open research problem.
% Addressing this research problem could help identifying effective usage of learned methods for rendering and storing 3D scenes in holographic displays.
% Our work explores this research problem with a preliminary study that could help guide more advanced experiments in the future.
Achieving robust compression remains an open challenge; our study guides future work on efficient 3D holographic rendering/storage.
\newline

\begin{table}[h!]
    \centering
    \caption{Patch based hologram quality comparison between \vanillaMLP, \FILMSIREN, and \SIREN.}
    \label{tbl:comparison}
    \begin{tabular}{@{}lccc@{}}
        \toprule
        \multicolumn{4}{@{}c@{}}{\textbf{\vanillaMLP}} \\
        \cmidrule{1-4}
        Patch size & PSNR $\pm$ Std. & Params & Comp. Ratio\\
        \cmidrule(r){1-1} \cmidrule(lr){2-2} \cmidrule(lr){3-3} \cmidrule(lr){4-4}
        
        $3\times 64\times64$ & 40.06 $\pm$ 2.73 & 5,059 & 41\% \\
        $\mathbf{3 \times 96\times96}$ & $\mathbf{41.50} \pm \mathbf{2.91}$ & $\mathbf{11,139}$ & $\mathbf{40\%}$ \\
        $3\times 128\times128$ & 39.88 $\pm$ 2.05 & 19,459 & 40\% \\
        $3 \times 160\times 160$ & 40.71 $\pm$ 1.89 & 31,939 & 41\% \\
        \midrule
        
        \multicolumn{4}{@{}c@{}}{\textbf{\FILMSIREN}} \\
        \cmidrule{1-4}
        Patch size & PSNR $\pm$ Std. & Params & Comp. Ratio\\
        \cmidrule(r){1-1} \cmidrule(lr){2-2} \cmidrule(lr){3-3} \cmidrule(lr){4-4}
        $\mathbf{3 \times 64\times 64}$ & $\mathbf{40.92} \pm \mathbf{2.91}$ & $\mathbf{4,869}$ & $\mathbf{40\%}$ \\
        $3 \times 96\times96$ & 40.68 $\pm$ 2.58 & 10,755 & 39\% \\ 
        $3 \times 128\times128$ & 39.70 $\pm$ 3.18 & 19,137 & 39\% \\
        $3 \times 160\times160$ & 35.48 $\pm$ 2.93 & 30,357 & 40\% \\
        \midrule
        
        \multicolumn{4}{@{}c@{}}{\textbf{\SIREN}} \\
        \cmidrule{1-4}
        Patch size & PSNR $\pm$ Std. & Params & Comp. Ratio\\
        \cmidrule(r){1-1} \cmidrule(lr){2-2} \cmidrule(lr){3-3} \cmidrule(lr){4-4}
        \arrayrulecolor{acmgreen}
        \midrule[1.5pt]
        $\mathbf{3 \times 64\times 64}$ & $\mathbf{42.29 \pm 2.45}$  & $\mathbf{4,899}$ & $\mathbf{40\%}$\\
        % $\mathbf{3 \times 96\times 96}$ & $\mathbf{40.83 \pm 2.63}$ & $\mathbf{0.99}$ & $\mathbf{11,171}$ & $\mathbf{40\%}$\\
        \midrule[1.5pt]
        \arrayrulecolor{black}
        $3 \times 96\times96$ & 40.83 $\pm$ 2.63 & 11,171 & 40\% \\
        $3 \times 128\times128$ & 39.32 $\pm$ 3.08 & 19,491 & 40\% \\
        $3 \times 160\times160$ & 37.51 $\pm$ 4.88 & 31,971 & 41\% \\
        \bottomrule
    \end{tabular}
\end{table}

\bibliographystyle{ACM-Reference-Format}
\bibliography{references}

\end{document}